\documentclass[aps,prb,superscriptaddress,reprint,twocolumn,longbibliography]{revtex4-1}

\usepackage{graphicx}
\usepackage{caption}
\usepackage{amssymb}
\usepackage{amsmath}
\usepackage{bm}
\newcommand{\vect}[1]{\bm{#1}}
\newcommand{\mat}[1]{\mathbf{#1}}
\newcommand{\Prob}{\mathbf{P}}

\DeclareCaptionLabelSeparator{bar}{\hspace{1ex}$|$}
\begin{document}
\title{
Transient fractality as a mechanism for emergent irreversibility in chaotic Hamiltonian dynamics}
\author{Y\^uto Murashita}
\author{Naoto Kura}
\affiliation{Department of Physics, University of Tokyo, 7-3-1 Hongo, Bunkyo-ku, Tokyo, 113-0033, Japan}
\author{Masahito Ueda}
\affiliation{Department of Physics, University of Tokyo, 7-3-1 Hongo, Bunkyo-ku, Tokyo, 113-0033, Japan}
\affiliation{RIKEN Center for Emergent Matter Science (CEMS), Wako, Saitama, 351-0198, Japan}
\date{\today}
\maketitle

{\bf
Understanding irreversibility in macrophysics from reversible microphysics has been the holy grail in statistical physics ever since the mid-19th century \cite{Boltzmann1872,Gibbs1875,Gibbs1902}.
Here the central question concerns the arrow of time\cite{Lebowitz1993}, which boils down to deriving macroscopic emergent irreversibility from microscopic reversible equations of motion.
As suggested by Boltzmann\cite{Boltzmann1877}, this irreversibility amounts to improbability (rather than impossibility) of the second-law-violating events.
Later studies\cite{HolianHooverPosch1987,GallavottiCohen1995} suggest that this improbability arises from a fractal attractor which is dynamically generated in phase space in reversible dissipative systems.
However, the same mechanism seems inapplicable to reversible conservative systems, since a zero-volume fractal attractor is incompatible with the nonzero phase-space volume, which is a constant of motion due to the Liouville theorem\cite{Gibbs1902}.
Here we demonstrate that in a Hamiltonian system the fractal scaling emerges transiently over an intermediate length scale.
Notably, this transient fractality is unveiled by invoking the Loschmidt demon\cite{Loschmidt1876} with an imperfect accuracy.
Moreover, we show that irreversibility from the fractality can be evaluated by means of information theory\cite{ErvenHarremoes2014} and the fluctuation theorem\cite{Seifert2012}.
The fractality provides a unified understanding of emergent irreversibility over an intermediate time scale regardless of whether the underlying reversible dynamics is dissipative or conservative.
}

In 1872, Boltzmann derived the $H$-theorem, which claims that the entropy is a nondecreasing function of time, and concluded that the state monotonically approaches equilibrium regardless of the initial state\cite{Boltzmann1872}.
However, his proof is based on the assumption of molecular chaos, where collisions of molecules are assumed to be memoryless.
Loschmidt refuted Boltzmann's idea by posing the irreversibility paradox that bears his name\cite{Loschmidt1876}:
under the laws of reversible mechanics, any phase-space trajectory can be traced back if we start from the final state with the reversed velocity.
Therefore, if we have a trajectory with positive entropy production, we should have its time-reversed trajectory with negative entropy production.
Hence, monotonicity of entropy is incompatible with microscopic reversible dynamics.
In response to Loschmidt's observation, Boltzmann argued that those states that result in positive entropy production appear much more frequently than those that lead to negative entropy production\cite{Boltzmann1877}.

This Boltzmann's insight was far ahead of the time -- it took a century before it was numerically verified.
By studying a dissipative system obeying the time-reversible Nos\'e-Hoover equations of motion \cite{Nose1984,Hoover1985}, Holian et al.\cite{HolianHooverPosch1987} observed that states in phase space collapse onto a fractal attractor and argued that one should sample a state precisely on the zero-volume fractal to cause a permanent violation of the second law.
This implies that trajectories with negative entropy production cannot be observed, not because of the violation of the equations of motion but because of the vanishing probability to sample the trajectories \cite{HolianHooverPosch1987,HooverHoover2012}.
It is illuminating to reconsider their discussion in view of the steady-state fluctuation theorem\cite{EvansCohenMorriss1993,GallavottiCohen1995,GallavottiCohen1995-2}.
According to this theorem, the probability of negative entropy production $P(-\sigma)$ is suppressed compared with that of the sign-reversed counterpart $P(\sigma)$ by an exponential factor: $P(-\sigma)/P(\sigma)\simeq e^{-\sigma}$.  
Therefore, the probability of second-law-violating events vanishes in the long run because of the time-extensive nature of entropy production.
It is noteworthy that fractal structures again play key roles in the derivation of the steady-state fluctuation theorem\cite{GallavottiCohen1995-2,GallavottiBook}.

Thus, the Loschmidt paradox in reversible dissipative systems has been settled on the basis of the fractality in phase space.
Here, the fractality arises because the equations of motion explicitly violate the Liouville theorem and therefore the phase-space volume contracts to zero.
Thus, the very original paradox posed in a Hamiltonian system remains unsolved.

In the present paper, we reveal that a hitherto unnoticed type of fractality arises transiently in a chaotic Hamiltonian system.
The Liouville theorem precludes infinitesimally structured fractal with zero volume but allows finite but exponentially small (e.g. sub-atomic-scale) fractal structures to emerge.
It is this type of transient fractality that yields emergent irreversibility in chaotic Hamiltonian systems.

\begin{figure*}
\begin{center}
	\begin{minipage}{183truemm}
		\begin{minipage}{0.24\textwidth}
			\flushleft{\bf a}\vspace{-0.05\textwidth}\\
			\centering
			\includegraphics[width=0.8\textwidth]{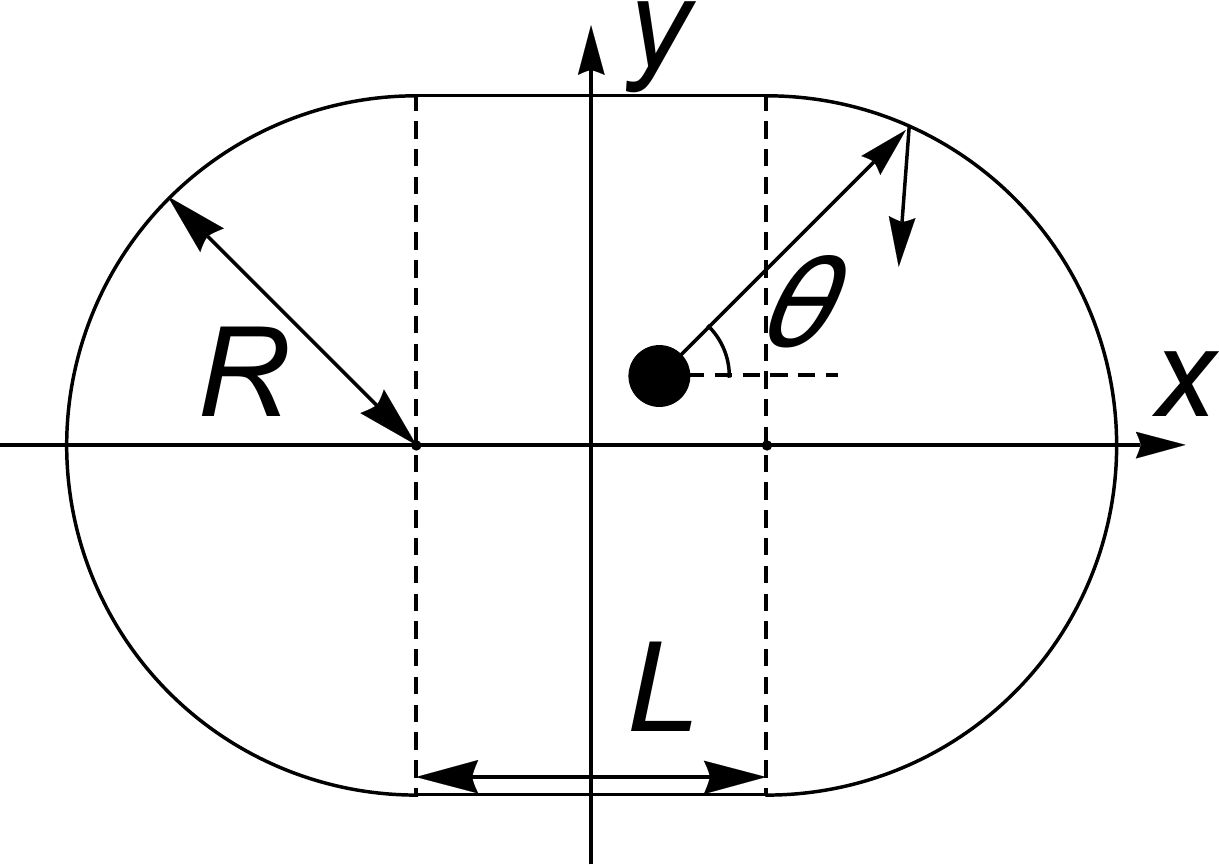}\vspace{-0.05\textwidth}\\
			\flushleft{\bf b}\vspace{-0.05\textwidth}\\
			\centering
			\includegraphics[width=0.9\textwidth]{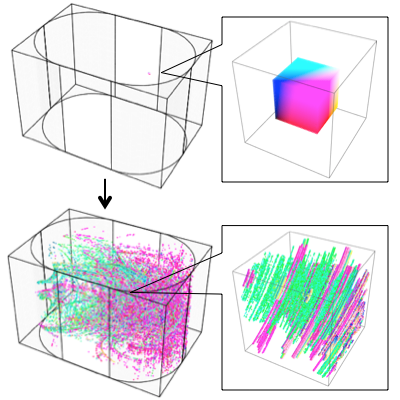}\\
			\flushleft{\bf c}\vspace{-0.05\textwidth}\\
			\centering
			\includegraphics[width=0.9\textwidth]{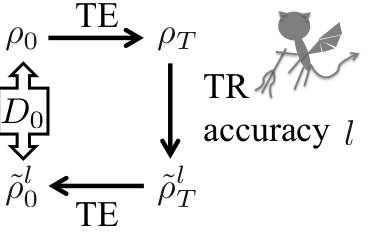}
		\end{minipage}
		\begin{minipage}{0.7\textwidth}
			\flushleft{\bf d}\hspace{0.08\textwidth}
			$\rho_0$\hspace{0.22\textwidth}
			$\rho_T$\hspace{0.2\textwidth}
			$\mathcal{T}\tilde\rho_T^l$\hspace{0.20\textwidth}
			$\mathcal{T}\tilde\rho_0^l$\\
			\includegraphics[width=\textwidth]{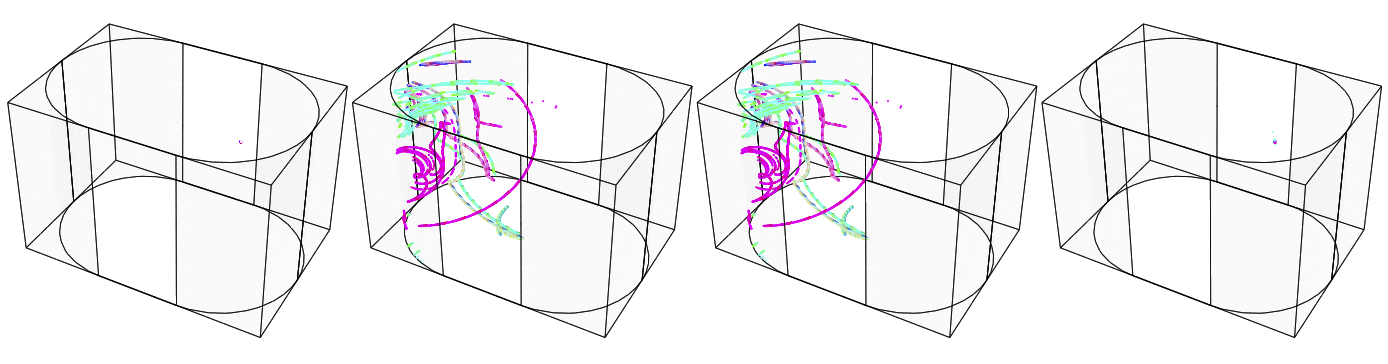}\\
			\flushleft\vspace{-0.09\textwidth}$T=20$\vspace{0.02\textwidth}
			\includegraphics[width=\textwidth]{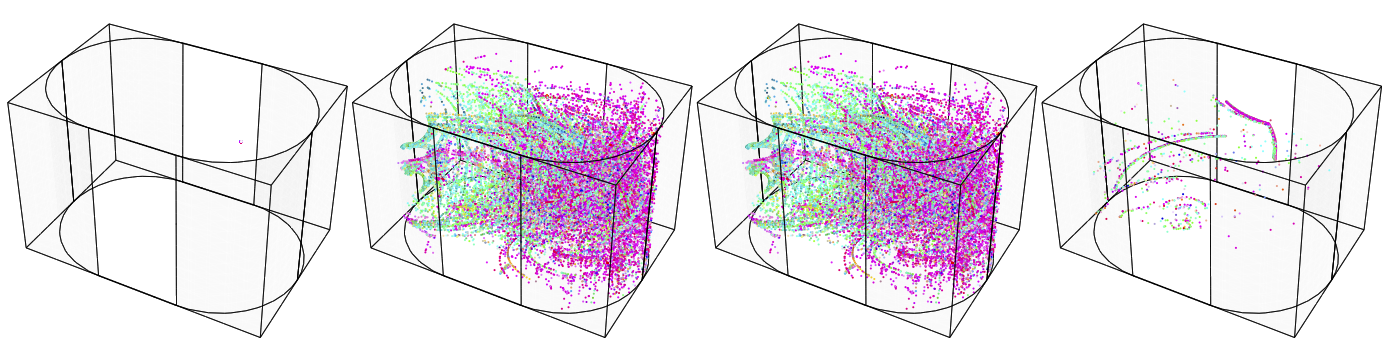}\\
			\vspace{-0.06\textwidth}$T=30$\vspace{0.02\textwidth}
			\includegraphics[width=\textwidth]{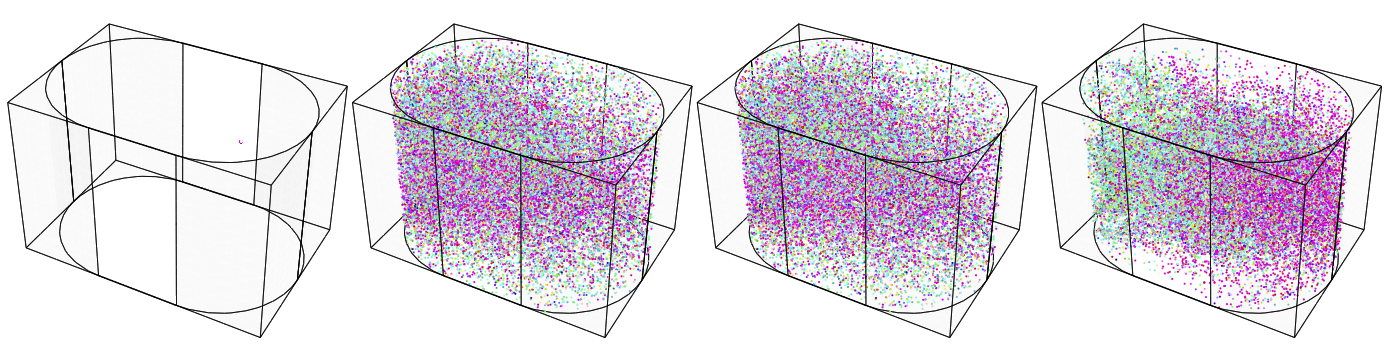}\\
			\vspace{-0.06\textwidth}$T=40$\vspace{0.02\textwidth}
		\end{minipage}
	\end{minipage}
	\caption{\label{fig:LoschmidtDemon}
		{\bf Time reversal test by an imperfect Loschmidt demon in a Bunimovich billiard.}
		{\bf a,}~Geometry of the Bunimovich stadium. A particle ballistically moves and elastically bounces on the boundary.
		{\bf b,}~Foliation in phase space.
		The phase-space structure evolves from a nonequilibrium state (a uniform distribution in a small cube) into a foliated structure, a typical feature of fractals.
		{\bf c,}~Schematic illustration of the time reversal test.
		The initial state $\rho_0$ undergoes time evolution (TE) over time $T$ to reach the final state $\rho_T$.
		Loschmidt's demon conducts the time reversal (TR) operation with accuracy $l$  and prepares the state $\tilde\rho_T^l$,
		which then evolves over time $T$ and ends up with the state $\tilde\rho_0^l$.
		Finally, we evaluate the R\'enyi-0 divergence $D_0(\rho_0||\mathcal{T}\tilde\rho^l_0)=D_0(\rho_T||\mathcal{C}_l[\rho_T])$, which reflects fractal vulnerability of $\rho_T$ against small perturbations.
		{\bf d,}~Time reversal test by the Loschmidt demon.
		When the evolution time $T$ is 20, $\mathcal{T}\tilde{\rho}^l_0$ is almost identical to the initial state $\rho_0$.
		However, as $T$ increases further, $\mathcal{T}\tilde\rho^l_0$ deviates from $\rho_0$ more significantly, which indicates an increase of the R\'enyi-0 divergence between them.
		Although $\rho_T$ and $\mathcal{T}\tilde\rho^l_T$ look identical, they actually differ for large $T$, which can be measured by the difference between $\rho_0$ and $\mathcal{T}\tilde\rho^l_0$.
		This points to the fact that the direct evaluation of $D_0(\rho_T||\mathcal{T}\tilde\rho^l_T)$ is extremely challenging without the time reversal test.
		See the Supplementary Information for details of numerical simulations.
	}
\end{center}
\end{figure*}

To probe the fractality, we employ the standard fractal measure of the box-counting dimension $d_{\rm F}$.
Specifically, we observe how the phase-space object of interest, such as the probability distribution function $\rho$, is vulnerable under small perturbations with typical length scale $l$.
Let $d_{\rm E}$ be the dimension of the embedding space, namely, phase space, and $\mathcal{C}_l[\rho]$ denote the convolution of $\rho$ with the $d_{\rm E}$-dimensional isotropic Gaussian with the standard deviation $l$.
We can show that the probability that the state remains in the original support after the Gaussian convolution satisfies the scaling law as $\int_{\rho(\gamma)>0} \mathcal{C}_l[\rho](\gamma)d\gamma\sim(l_0/l)^{d_{\rm E}-d_{\rm F}}$, where $l_0$ is the minimal length scale of $\rho$.
Interestingly, the left-hand side can be rewritten by an information-theoretic measure called the R\'enyi-0 divergence \cite{ErvenHarremoes2014}, which is defined as $D_0(\rho||\mathcal{C}_l[\rho]):=-\ln\int_{\rho(\gamma)>0} \mathcal{C}_l[\rho](\gamma)d\gamma$, and the scaling law is translated as
\begin{equation}
\label{eq:ScalingLaw}
	D_0(\rho||\mathcal{C}_l[\rho])
	=
	d_{\rm C}\ln(l/l_0),
\end{equation}
where we define the fractal codimension by $d_{\rm C}:=d_{\rm E}-d_{\rm F}$.
Consequently, we can extract the fractal dimension of $\rho$ by observing the behaviors of the R\'enyi-0 divergence against the variations of the length scale $l$.
Although the scaling law (\ref{eq:ScalingLaw}) mathematically applies to the limit of $l\to 0$, we utilize it for sufficiently small $l(\neq 0)$ to probe fractality in a physical sense.
See the Supplementary Information for details.

\begin{figure}
\begin{center}
	\begin{minipage}{89truemm}
		\flushleft{\bf a}\hspace{0.47\textwidth}{\bf b}\\
		\includegraphics[width=0.47\textwidth]{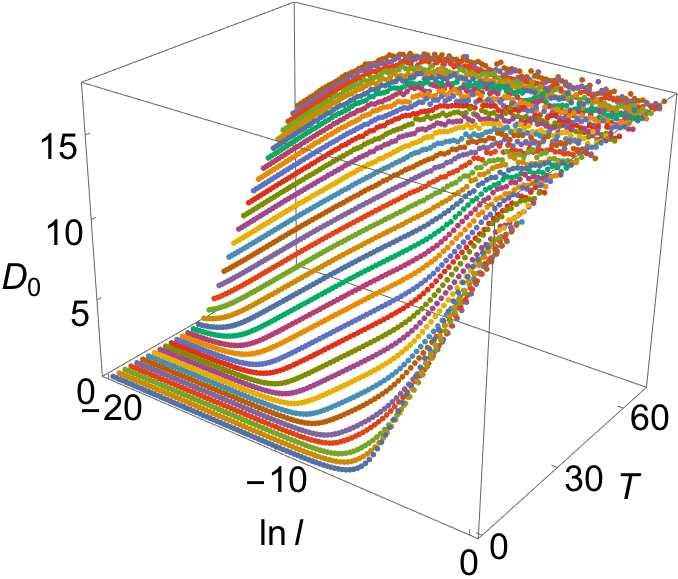}
		\includegraphics[width=0.51\textwidth]{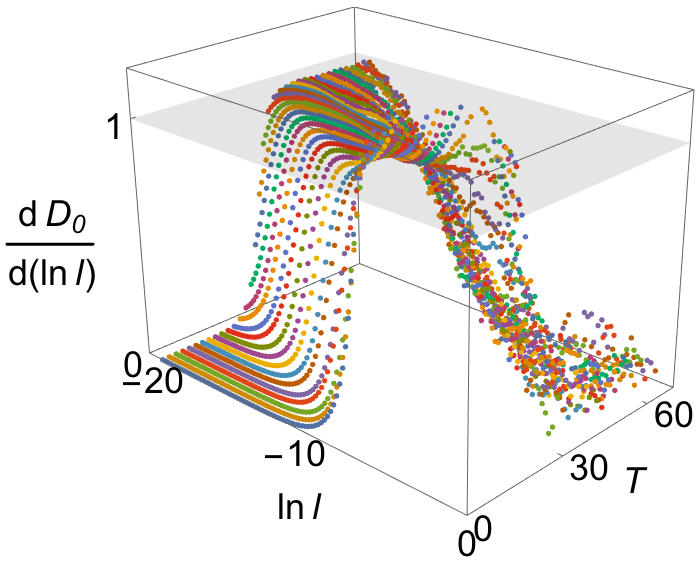}
	\end{minipage}
	\caption{\label{fig:renyi}
		{\bf Transient fractal structure in phase space disclosed by the R\'enyi-0 divergence.}
		{\bf a,}~R\'enyi-0 divergence $D_0$ against variations of the minimal length scale $l$ and the evolution time $T$.
		The linear dependence of $D_0$ on $\ln l$ in an intermediate length scale indicates that the phase-space structure satisfies the scaling law~(\ref{eq:ScalingLaw}) unique to fractality.
		{\bf b,}~Derivative of the R\'enyi-0 divergence $D_0$ with respect to $\ln l$.
		The plateau in the intermediate length scale corresponds to the transient region where the fractal scaling law applies.
		The value of the plateau is $d_{\rm C}=1.0$, which indicates that the fractal dimension is $d_{\rm F}=2.0$.
		The gray plane is an eyeguide with $d_{\rm C}=1$.
		See the Supplementary Information for details of numerical simulations.
	}
\end{center}
\end{figure}

As a chaotic Hamiltonian system, we consider an elastic billiard in the Bunimovich stadium\cite{Bunimovich1979} (see Fig.~\ref{fig:LoschmidtDemon}a).
The phase space of the Bunimovich billiard can be represented as a three-dimensional space ($d_{\rm E}=3$) due to the conservation of energy; the coordinate of the phase space consists of two positions ($x$ and $y$) and the direction of the velocity ($\theta$).
We use the Monte-Carlo method to study this system.
First of all, we sample a phase-space point $\gamma_0$ from the probability distribution function at the initial time $\rho_0$, which is prepared as a uniform distribution in a small cube.
Then, we let it evolve over time $T$ to obtain $\gamma_T$, which collectively constitutes the final state $\rho_T$.
To evaluate the fractality, we calculate the R\'enyi-0 divergence between $\rho_T$ and the perturbed state $\mathcal{C}_l[\rho_T]$.
To this aim, we add to $\gamma_T$ a Gaussian noise with the standard deviation $l$ and judge whether the resulting point $\gamma_T'$ belongs to the support of $\rho_T$, contributing to the integral in $D_0$.
Unfortunately, this judgement is impractical because $\rho_T$ has a complicated support with an exponentially elongated boundary (see Fig.~\ref{fig:LoschmidtDemon}b).

The imperfect Loschmidt demon plays a key role in evaluateing the R\'enyi-0 divergence (see Fig.~\ref{fig:LoschmidtDemon}c).
The demon invokes the protocol called the time-reversal test\cite{Shepelyansky1983,AdachiTodaIkeda1988}.
For each point $\gamma_T'$ in $\mathcal{C}_l[\rho_T]$, the demon inverts the velocity to obtain the corresponding point $\tilde\gamma_T=\mathcal{T}\gamma_T'$, which constitutes an ensemble $\tilde \rho^l_T = \mathcal{T}\mathcal{C}_l[\rho_T]$, where $\mathcal{T}$ represents the velocity reversal.
Then, the demon lets the point evolve over time $T$ to reach the point $\tilde \gamma_0$ and the corresponding ensemble $\tilde \rho^l_0$.
The fraction of $\tilde\rho^l_0$ in the original support of $\rho_0$ contributes to the integral in $D_0$.
We note that our protocol is similar to the Loschmidt echo\cite{JalabertPastawski2000}, which is used to characterize decoherence in classically chaotic quantum systems.
The difference is that the backward Hamiltonian (rather than the state $\rho_T$) is perturbed in the Loschmidt echo.

When $l$ is small, the difference between $\rho_T$ and $\mathcal{T}\tilde\rho^l_T$ cannot be discerned (see Fig.~\ref{fig:LoschmidtDemon}d).
However, after the reversed time evolution, the difference between $\rho_0$ and $\mathcal{T}\tilde\rho^l_0$ is manifested, which is quantitatively characterized by the R\'enyi-0 divergence $D_0$.
For $T=20$, the state $\mathcal{T}\tilde{\rho}_0^l$ returns to the initial state almost completely. However, as $T$ further increases, $\mathcal{T}\tilde\rho^l_0$ deviates from the initial state $\rho_0$ and expands over the entire phase space.
This difference results in a larger value of $D_0$, implying increasing vulnerability of the fractal structure of $\rho_T$.

Figure~\ref{fig:renyi}a plots the R\'enyi-0 divergence $D_0$ by changing the length scale $l$ and time $T$.
For intermediate values of $l$ and $T$, we can observe $D_0$ is on a plane.
This behavior indicates that $D_0$ satisfies the scaling law of a fractal in equation~(\ref{eq:ScalingLaw}).
In fact, by substituting $l_0\simeq w e^{-\Lambda T}$ into equation~(\ref{eq:ScalingLaw}), we obtain
\begin{equation}
	\label{D0plane}
	D_0 \simeq
	d_{\rm C} \ln l + d_{\rm C}\Lambda T + {\rm const.}
\end{equation}
To see the $l$-dependence more clearly, Fig.~\ref{fig:renyi}b gives the derivative of $D_0$ with respect to $\ln l$.
For a fixed $T$, a plateau can be observed in the intermediate regime of the length scale $l$, which is the evidence of the linear behavior of $D_0$ with respect to $\ln l$.
Moreover, since the value of the plateau corresponds to the fractal codimension $d_{\rm C}$, we find $d_{\rm C}=1.0$.
Therefore, we conclude that the structure with the fractal dimension $d_{\rm F}=2.0$ is dynamically generated in the intermediate time scale for a fixed $l$.
As a result, the proportionality factor for $T$ amounts to $\Lambda$, which coincides with the Kolmogorov-Sinai (KS) entropy $h_{\rm KS}$ of this system according to the Pesin formula~\cite{ZaslavskyBook}.
We note that a similar linear growth, i.e., $h_{\rm KS} T+{\rm const.}$, in an intermediate time was found for a coarse-grained Shannon entropy in chaotic conservative maps~\cite{LatoraBaranger1999}.
However, no mention about the fractality is made in ref.~\onlinecite{LatoraBaranger1999}.
In the Supplementary Information, we give a mathematical argument in support of this emergent fractality and a discussion on numerical accuracy.

We stress that this transient fractality is compatible with the Liouville theorem.
The Liouville theorem states that the phase-space volume is conserved.
Meanwhile, the mathematical fractal with infinitely fine structures has zero volume.
Therefore, the initial state with nonzero volume cannot evolve into an ideal mathematical fractal.
Consequently, fractal structures seem impossible in the Hamiltonian system.
However, we here consider the fractal scaling in a fixed length scale $l$.
By fixing $l$, the phase-space object with nonzero volume can satisfy the fractal scaling law as long as it has a finer structure than $l$.
Moreover, with $l$ fixed, the state generates finer and finer structures over time until it finally looks uniformly nonzero.
This is why fractality is transient in the Hamiltonian system with the Liouville theorem.
We note that this transient fractal is different from the so-called fat fractal\cite{UmbergerFarmer1985}, which has infinitely fine structures and positive volume at the same time.

\begin{figure*}
\begin{center}
\begin{minipage}{183mm}
	\begin{minipage}{0.5\textwidth}
		\flushleft{\bf a}\vspace{-0.05\textwidth}\\
		\hspace{0.03\textwidth}
		\includegraphics[width=0.9\textwidth]{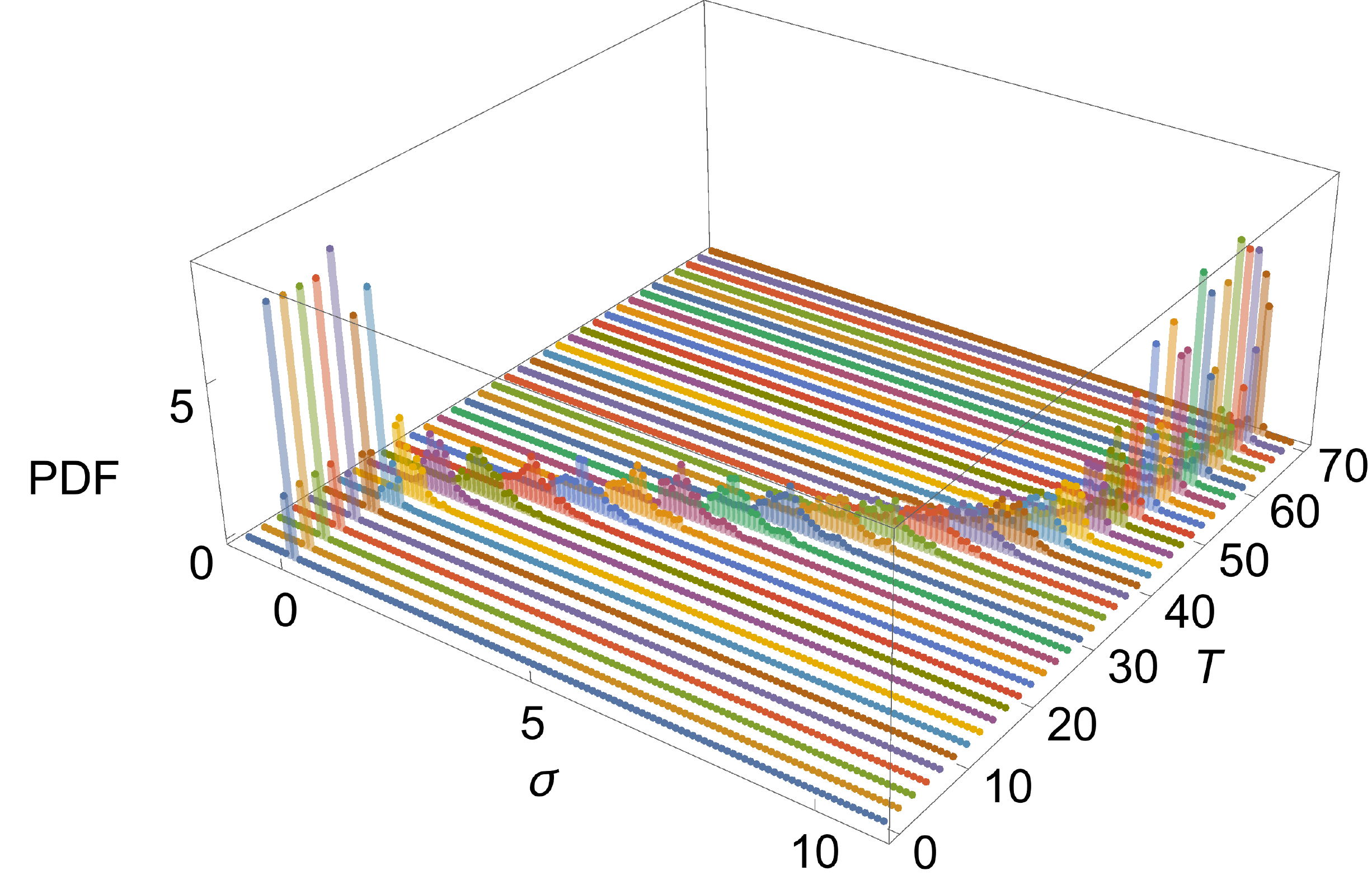}
	\end{minipage}
	\begin{minipage}{0.47\textwidth}
		\flushleft{\bf b}\hspace{0.47\textwidth}{\bf c}\vspace{-0.01\textwidth}\\
		\includegraphics[width=0.48\textwidth]{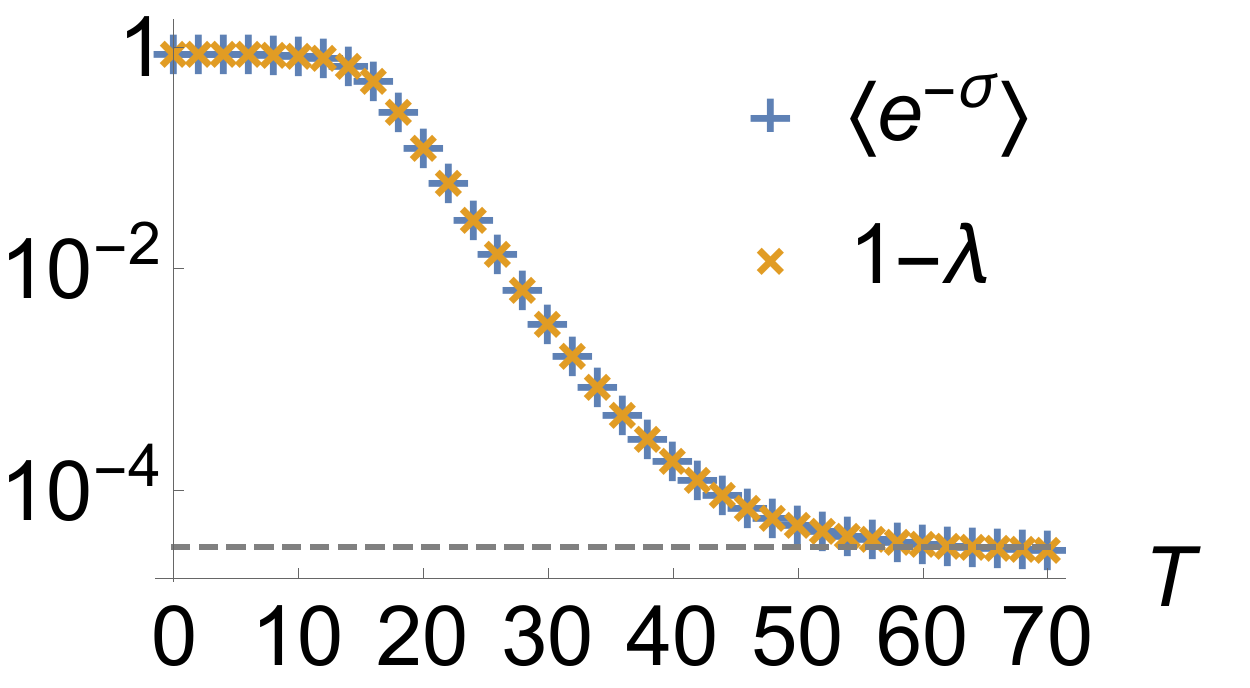}
		\includegraphics[width=0.5\textwidth]{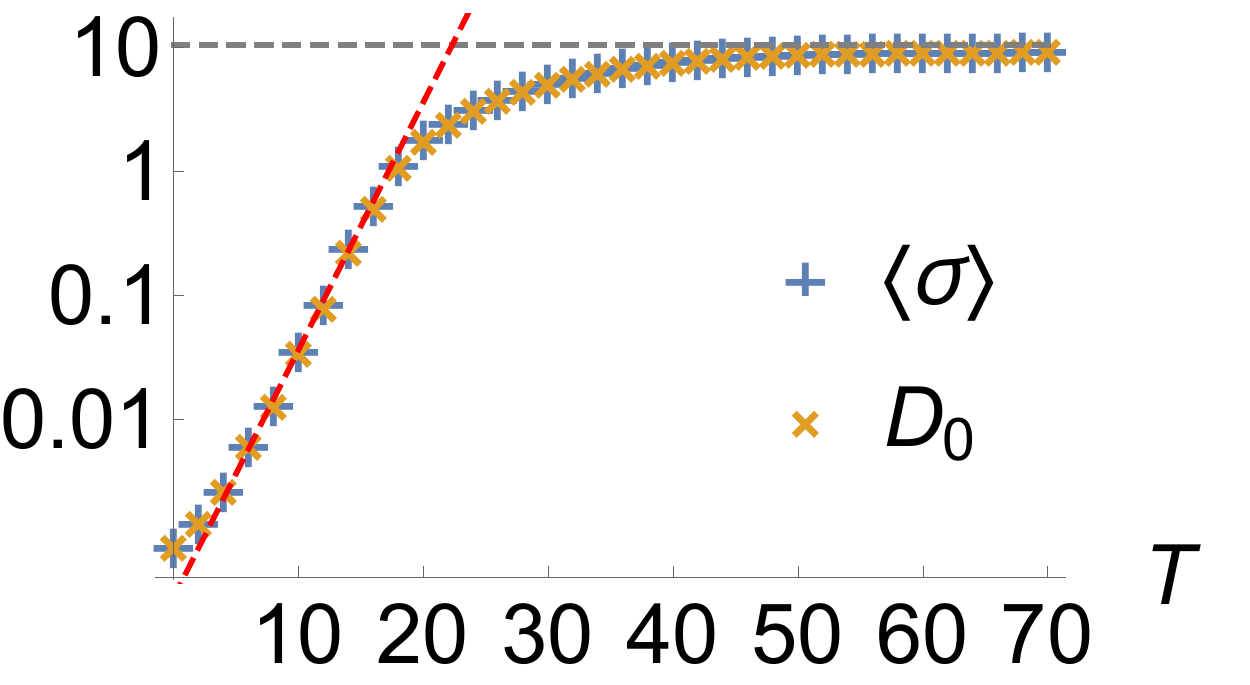}
		\flushleft{\bf d}\hspace{0.47\textwidth}{\bf e}\vspace{-0.03\textwidth}\\
		\hspace{0.01\textwidth}
		\includegraphics[width=0.47\textwidth]{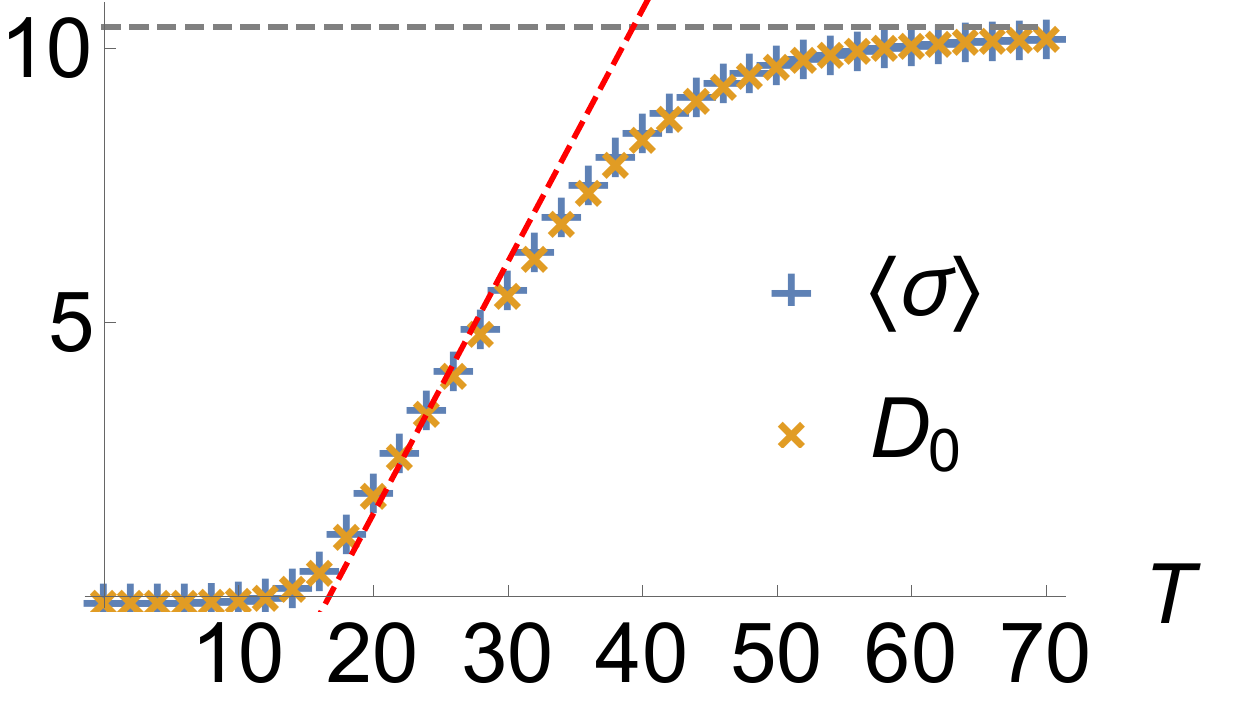}
		\hspace{0.01\textwidth}
		\includegraphics[width=0.47\textwidth]{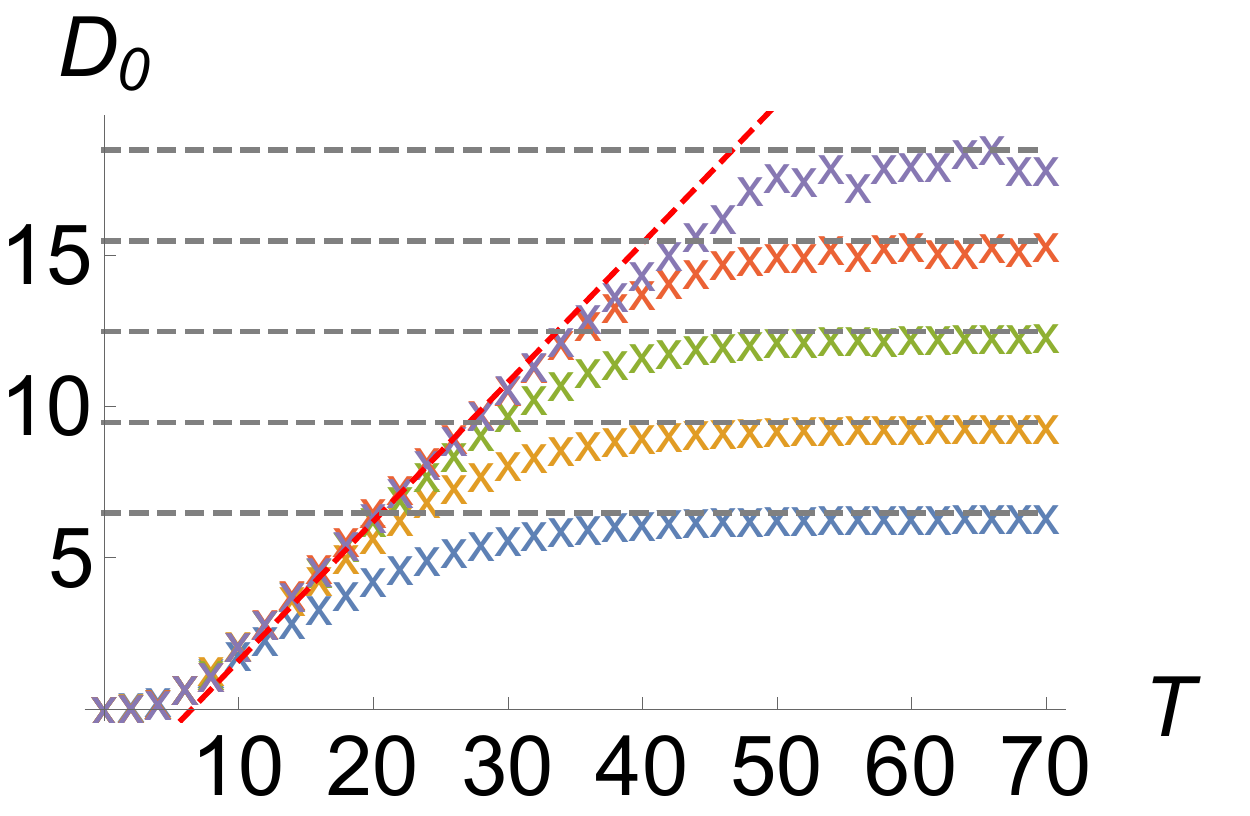}
	\end{minipage}
\end{minipage}
	\caption{\label{fig:ent}
		{\bf Empirical entropy production and the fluctuation theorem with absolute irreversibility.}
		{\bf a,}~Time evolution of the probability distribution function (PDF) of the empirical entropy production $\sigma$.
		The PDF shifts toward larger $\sigma$ over time, indicating stronger irreversibility in the time-reversal test.
		{\bf b,}~Verification of the fluctuation theorem with absolute irreversibility in equation~(\ref{eq:FT}).
		The exponential averages of the empirical entropy production $\langle e^{-\sigma}\rangle$ calculated from the PDF of $\sigma$ and the values of absolute irreversibility $\lambda$ are plotted against time.
		The excellent agreement between them indicates the validity of the fluctuation theorem with absolute irreversibility.
		The gray dotted line indicates the saturated value due to a finite-size effect.
		{\bf c,} Bound by the second-law-like inequality~(\ref{eq:ineq}) shown in the logarithmic scale.
		The red dashed line represents exponential growth with respect to time.
		{\bf d,} Same data in the linear scale.
		The red dashed line shows the linear growth in equation~(\ref{D0plane}), which emerges after the exponential growth.
		{\bf e,} Saturation and finite-size effects.
		By reducing the initial phase-space volume (from below to top), we can increase the saturated value due to the finite-size effect.
		See the Supplementary Information for details of numerical simulations.
	}
\end{center}
\end{figure*}

We consider irreversibility based on a variation of the finite-time fluctuation theorem\cite{Jarzynski1997,Crooks2000,HatanoSasa2001}.
The  detailed fluctuation theorem compares the path probability in the original physical dynamics $P[\Gamma]$ with that in its appropriately chosen dual dynamics $\tilde P[\Gamma]$\cite{EspositoBroeck2010,Seifert2012}.
The ratio of these probabilities corresponds to the entropy production: $\tilde P[\Gamma]/P[\Gamma]=e^{-\sigma[\Gamma]}$.
We here take the dual dynamics to be the imperfect time-reversed process $\tilde \rho^l_T \to \tilde \rho^l_0$ (see Fig.~\ref{fig:LoschmidtDemon}c) and define an empirical entropy production by $\sigma=-\ln(\tilde P[\Gamma]/P[\Gamma])$.
Since the dynamics is deterministic in our case, the path probability is determined only by the initial point and therefore we obtain
$
	\sigma
	=
	-\ln [{\mathcal{T}\tilde\rho^l_0(\gamma_0)}/{\rho_0(\gamma_0)}].
$
Note that the ensemble average $\langle\sigma\rangle$ is not the thermodynamic entropy production but an information-theoretic measure called the Kullback-Leibler divergence\cite{ErvenHarremoes2014}:
$
	D_1(\rho_0||\mathcal{T}\tilde\rho_0^l)
	:=
	\langle \ln[\rho_0(\gamma)/\mathcal{T}\tilde\rho^l_0(\gamma)]\rangle
	:=
	\int\rho_0(\gamma)\ln[\rho_0(\gamma)/\mathcal{T}\tilde\rho^l_0(\gamma)]d\gamma
$.
Since $D_1(\rho_0||\mathcal{T}\tilde\rho^l_0)=D_1(\rho_T||\mathcal{T}\tilde\rho^l_T)$, it gives a (quasi-)distance between the fractal state $\rho_T$ and the state $\mathcal{T}\tilde\rho^l_T=\mathcal{C}_l[\rho_T]$ that one observes with nonzero accuracy $l$.

Notably, the empirical entropy production does not satisfy the conventional Jarzynski-type equality\cite{Jarzynski1997}: $\langle e^{-\sigma}\rangle\neq1$.
This is because $\sigma$ is divergent when $\gamma_0$ is outside the support of $\rho_0$ as $\sigma=-\ln[\mathcal{T}\tilde\rho^l_0(\gamma_0)/0]$.
This phenomenon with divergent entropy production should be distinguished from ordinary irreversibility and referred to as absolute irreversibility\cite{MurashitaFunoUeda2014}.
In the presence of absolute irreversibility, the Jarzynski-type equality should be modified\cite{MurashitaFunoUeda2014} as
\begin{equation}
\label{eq:FT}
	\langle e^{-\sigma} \rangle
	=
	1-\lambda,
\end{equation}
where $\lambda$ is the probability of the singular part of $\tilde\rho^l_0$ with respect to $\rho_0$ in the measure-theoretic sense (see e.g. refs.~\onlinecite{Halmos1974,Bartle1995}).
In this case, since $\lambda=\int_{\rho(\gamma)=0}\mathcal{T}\tilde\rho^l_0(\gamma)d\gamma$, we have $1-\lambda=e^{-D_0}$.
By applying the Jensen inequality to equation~(\ref{eq:FT}), we obtain
\begin{equation}
\label{eq:ineq}
	\langle \sigma \rangle
	=
	D_1(\rho_T||\mathcal{C}_l[\rho_T])
	\ge
	-\ln(1-\lambda)
	=
	D_0(\rho_T||\mathcal{C}_l[\rho_T]).
\end{equation}
We note that the magnitude relation between $D_0$ and $D_1$ is well-known in information theory~\cite{ErvenHarremoes2014}.
In Fig.~\ref{fig:ent}a, we numerically calculate the probability distribution function of the empirical entropy production $\sigma$, and verify the fluctuation theorem with absolute irreversibility in Fig.~\ref{fig:ent}b.
Figures 3c and 3d compare the empirical entropy production $\langle\sigma\rangle$ with the R\'enyi-0 divergence $D_0=-\ln(1-\lambda)$ in logarithmic and linear scales, respectively.
It can be seen that $\langle \sigma \rangle$ is bounded from below by $D_0$.

Moreover, we can see that three different time regimes exist.
Initially, $D_0$ exponentially increases as $e^{\Lambda T}$, where $\Lambda$ is the positive Lyapunov exponent (see Fig.~\ref{fig:ent}c).
Then, in the following intermediate time regime, $D_0$ grows linearly in time with the proportionality factor of $h_{\rm KS}$ as shown in Fig.~\ref{fig:ent}d (see equation~(\ref{D0plane})).
Finally, $D_0$ gets saturated to be the value determined by the ratio of the initial-state volume to the entire phase-space volume.
We note that the behaviors in these three time scales are analogous to the ones of the coarse-grained Shannon entropy~\cite{LatoraBaranger1999}.
We can postpone the occurrence of this finite-size effect to a later time by reducing the initial volume and enlarge the time domain for the transient fractal as shown in Fig.~\ref{fig:ent}e.
The linear growth in the transient time scale is reminiscent of the constant entropy production in a steady state of a dissipative system.

As discussed in the Supplementary Information, all of these arguments together with mathematical discussions lead us to conjecture that a generic chaotic Hamiltonian system with the spatial dimension $d$ has transient fractality with $d_{\rm F}=d$ and exhibits a transient linear growth of the empirical entropy with the rate identical to the KS entropy as long as the finite-size effect is negligible.
Thus, we suggest that the transient fractality constitutes a universal mechanism of emergent irreversible phenomena in generic reversible dynamics.

We have demonstrated that transient fractality emerges in the Bunimovich billiard even though it obeys the Liouville theorem.
An imperfect Loschmidt demon plays a crucial role in evaluating the R\'enyi-0 divergence characterizing fractality.
By regarding the process generated by the demon as the dual process for the fluctuation theorem, we have evaluated the empirical entropy production $\sigma$ and verified that it satisfies the fluctuation theorem with absolute irreversibility.
The average of $\sigma$ gives how the state $\rho_T$ is different from the state $\mathcal{T}\tilde\rho^l_T=\mathcal{C}_l[\rho_T]$ that the demon or an observer perceives by probing $\rho_T$ with nonzero accuracy $l$.
Therefore, the state $\mathcal{T}\tilde \rho^l_0$ that we retrodict to be the initial state from our observation at time $T$ becomes more distinct from the initial state $\rho_0$.
Thus, the averaged empirical entropy production $\langle\sigma\rangle$ captures this irreversible informational loss caused by the vulnerability of the transient fractal.
We expect that this information-theoretic irreversibility is related to the thermodynamic irreversibility.
In particular, it is worth investigating the relation between the empirical and thermodynamic entropy productions.
An extension of our classical study to a quantum regime should merit further study because the natural length scale determined by the Planck constant $\hbar$ should provide a natural cutoff in addition to $l$ and there should be an interplay between them.
Moreover, our fractal picture of irreversibility might give a unified framework to describe information loss in equilibriation  regardless of whether a system is isolated or open, and classical or quantum.

\bibliography{reference}

\noindent
{\bf Acknowledgement} This work was supported by a Grant-in-Aid for Scientific Research on Innovative Areas ``Topological Materials Science'' (KAKENHI Grant No. JP15H05855) and the Photon Frontier Network Program from MEXT of Japan.
Y.M. was supported by the Japan Society for the Promotion of Science (JSPS) through the Program for Leading Graduate School (MERIT) and a JSPS Fellowship (Grant No. JP15J00410).
N.K. was supported by Advanced Leading Graduate Course for Photon Science (ALPS) of JSPS.

\noindent
{\bf Author Contributions}
Y.M. designed the numerical simulations and developed the physical aspect of the theory.
N.K. provided mathematical supports.
Y.M. and N.K. wrote the manuscript of the main text and the Supplementary Information, respectively, with feedback from the others.
All work was done under the supervision of M.U.

\widetext
\pagebreak
\begin{center}
{\large Supplementary Information for ``Transient fractality as a mechanism for emergent irreversibility in chaotic Hamiltonian dynamics''}
\end{center}

\renewcommand{\theequation}{S\arabic{equation}}
\renewcommand{\thefigure}{S\arabic{figure}}
\renewcommand{\bibnumfmt}[1]{[S#1]}
\setcounter{equation}{0}
\setcounter{figure}{0}

\section{Fractal dimension and the R\'enyi-0 divergence}
 In our main text, the fractal dimension $d_\mathrm{F}$ is determined from the scaling law of the R\'enyi-0 divergence:
 \begin{equation}
  \begin{aligned}
   D_0\bigl(\rho\Vert \mathcal{C}_l[\rho]\bigr) &= d_\mathrm{C}\ln (l/l_0), \\
   d_\mathrm{C} &= d_\mathrm{E} - d_\mathrm{F},
  \end{aligned}\label{TransientFractality} 
\end{equation}
 where $d_\mathrm{E}$ is the dimension of the embedding space and $l_0$ is the smallest length scale of the fractal.
 From a viewpoint of rigorous mathematics, fractals should have an infinitesimal structure ($l_0\to 0$).
 However, from a viewpoint of physics, any fractal generated by a finite number of operations should have a nonzero (albeit extremely small) $l_0$.
We refer to fractals with an infinitesimal structure as mathematical fractals, and those with a nonzero minimal length scale as physical fractals.
 In this section, we verify equation~(\ref{TransientFractality}) for physical fractals on the basis of some natural assumptions.

 First, we review how the fractal dimensions are mathematically defined.
 Let $S$ be a subset in the $d_\mathrm{E}$-dimensional Euclidean space.
 For a fixed $a>0$, we consider a hypercubic lattice with spacing $a$, which partitions the space into $d_\mathrm{E}$-dimensional hypercubes (or ``boxes'') of volume $a^{d_\mathrm{E}}$.
 We denote by $N(a)$ the number of boxes that have nonvanishing overlap with $S$.
 The \emph{box-counting dimension} $d_\mathrm{box}$ is then defined as the limit of
 \begin{equation}
  d_\mathrm{box} = -\lim_{a \to +0} \frac{\ln N(a)}{\ln a}.
   \label{BoxCountingDim}
 \end{equation}
 
 Meanwhile, given the probability distribution $\pi$ supported on $S$, it is possible to obtain generalized fractal dimensions \cite{hentschel83} including the \emph{correlation dimension}.
 For that purpose, we cover the fractal $S$ with $N(a)$ boxes, which are labeled with integers from $1$ to $N(a)$.  
 Let $\pi_i(a)$ denote the probability with which a sample taken from $\pi$ falls on the $i$-th box.
 The correlation dimension $d_\mathrm{cor}$ is defined as
 \begin{equation}
  d_\mathrm{cor} = \lim_{a \to +0} \frac{\ln C(a)}{\ln a},
   \label{CorrelationDim}
 \end{equation}
 where $C(a)$ is 
 \begin{equation}
  C(a) = \sum_{i=1}^{N(a)} \bigl(\pi_i(a)\bigr)^2.
   \label{CorrelationByBox}
 \end{equation}
 Although $d_\mathrm{box}=d_\mathrm{cor}$ does not generally hold, this equality is satisfied in many practical cases.
 For instance, we may assume $\pi$ to be uniform in the sense that the relation
 \begin{equation}
  \pi_i(a) = \frac{1}{N(a)} + o\biggl(\frac{1}{N(a)}\biggr) \qquad (i=1, \ldots, N(a))
 \end{equation}
 holds for all $a>0$.
 In this case, we obtain $C(a)=1/N(a) + o\bigl(1/N(a)\bigr)$ and the two fractal dimensions coincide with each other.
 As we discuss later, the assumption of uniformity is consistent with the time reversal test in the main text.
 Henceforth, we assume that $\pi$ has a unique fractal dimension $d_\mathrm{F}=d_\mathrm{box}=d_\mathrm{cor}$.
 
 It is also known that the function $C(a)$ in equation~(\ref{CorrelationDim}) can be replaced by the integral:
 \begin{equation}
  C'(a) = \iint \exp \biggl(-\frac{\lVert \gamma'-\gamma \rVert^2}{2a^2}\biggr) \pi(d\gamma)\pi(d\gamma').
   \label{CorrelationByGauss}
 \end{equation}
 Both $C(a)$ and $C'(a)$ roughly estimate the probability with two samples $\gamma, \gamma'$ independently chosen from $\pi$ fall within the distance of $a$.
 
 Now, we consider a physical fractal that reduces to a mathematical fractal $S$ as long as we consider a length scale larger than $l_0$.
 We define $S'$ to be the union of $N(l_0)$ boxes covering $S$.
 The length scale $l_0$ describes the smallest structure in the physical fractal, with which $S'$ can be regarded to be the same as the mathematical fractal $S$.

 Furthermore, we consider a uniform probability distribution over $S'$.
 We note that the uniformity of the distribution also holds for the final state $\rho_T$ in the time reversal test, since the initial state is uniform and the dynamics is volume-preserving.
 The probability density function $\rho$ can be written as
 \begin{equation}
  \rho(\gamma) = \frac{\mathbf{1}_{S'}(\gamma)}{l_0^{d_\mathrm{E}}N(l_0)},
   \label{UniformPDF}
 \end{equation}
 where $\mathbf{1}_{S'}$ is the indicator function of $S'$, which returns $1$ for $\gamma\in S'$ and $0$ for $\gamma\notin S'$.
 
 The R\'enyi-0 divergence $D_0\bigl(\rho \big\Vert \mathcal{C}_l[\rho]\bigr)$ can be computed as
 \begin{align}
  D_0\bigl(\rho \big\Vert \mathcal{C}_l[\rho]\bigr)
  &= -\ln \int_{\rho(\gamma)>0} \mathcal{C}_l[\rho](\gamma)d\gamma
   = -\ln \int \mathcal{C}_l[\rho](\gamma)\mathbf{1}_{S'}(\gamma) d\gamma \\ 
  &= -\ln \iint \frac{1}{\bigl(\sqrt{2\pi}l\bigr)^{d_\mathrm{E}}}\exp \biggl(-\frac{\lVert \gamma'-\gamma \rVert^2}{2l^2}\biggr)
  \rho(\gamma')\mathbf{1}_{S'}(\gamma) d\gamma d\gamma'.
 \end{align}
 We further use \eqref{UniformPDF} to obtain
 \begin{align}
  D_0\bigl(\rho \big\Vert \mathcal{C}_l[\rho]\bigr)
  &= -\ln \iint \frac{l_0^{d_\mathrm{E}} N(l_0)}{\bigl(\sqrt{2\pi}l\bigr)^{d_\mathrm{E}}}
  \exp \biggl(-\frac{\lVert \gamma'-\gamma \rVert^2}{2a^2}\biggr)
  \rho(\gamma)\rho(\gamma') d\gamma d\gamma'\\
  &= d_\mathrm{E}\ln \biggl(\sqrt{2\pi}\frac{l}{l_0}\biggr) - \ln N(l_0)
  - \ln \iint \exp \biggl(-\frac{\lVert \gamma'-\gamma \rVert^2}{2l^2}\biggr) \rho(\gamma)\rho(\gamma') d\gamma d\gamma'.
  \label{ConstantAndIntegral}
 \end{align}
 The integral in \eqref{ConstantAndIntegral} coincides with $C'(l)$ if we replace the measure $\rho(\gamma)d\gamma$ by $\pi(d\gamma)$.
 This replacement is justified when $l$ is sufficiently larger than $l_0$, because the difference between $\rho(\gamma)d\gamma$ and $\pi(d\gamma)$ comes into play only in a length scale below $l_0$.
 Therefore, for $1\gg l\gg l_0$ the R\'enyi-0 divergence is evaluated as
 \begin{align}
  D_0\bigl(\rho \big\Vert \mathcal{C}_l[\rho]\bigr)
  &\approx d_\mathrm{E}\ln \biggl(\sqrt{2\pi}\frac{l}{l_0}\biggr) - \ln N(l_0) - \ln C'(l) \\
  &\approx d_\mathrm{E}\ln (l/l_0) + d_\mathrm{box}\ln l - d_\mathrm{cor}\ln l_0.
 \end{align}
 Here the approximations $\ln N(l_0)\approx -d_\mathrm{box}\ln l_0$ and $\ln C(l_0)\approx d_\mathrm{cor}\ln l_0$ follow from the definition of the fractal dimensions \eqref{BoxCountingDim} and \eqref{CorrelationDim}.
 We obtain the desired relation \eqref{TransientFractality} by substituting $d_\mathrm{F}=d_\mathrm{box}=d_\mathrm{cor}$ in the last equation.

\section{Conjecture about the fractal dimension in a Hamiltonian system with a generic dimension}
In this section, we discuss the transient fractality in a generic isolated time-reversal-invariant Hamiltonian system.  By a chaos-theoretic approach with differential geometry, we claim that a transient fractal of dimension $d_\mathrm{F}=d$ should emerge, where $d$ is the spatial dimension of the Hamiltonian system.

We denote by $M$ the phase space of the Hamiltonian system.
Under the constraint of energy conservation, $M$ has the dimension of $d_\mathrm{E}=2d-1$ and can therefore be represented by some local coordinates $(x_1, \dotsc, x_{2d-1})$.
In addition, we assume that the support $S_0$ of the initial state $\rho_0$ be sufficiently small.  More specifically, the spread of $S_0$ must be comparable to $w$ ($\ll 1$) in all directions.  We exclude the case in which the extent of $S_0$ in one direction is by far larger than that in the other directions.  Cubes and spheres satisfy this condition.

Following the procedure of the time reversal test, we sample a phase-space point $\gamma_0$ from the initial state $\rho_0$, which evolves into $\gamma_T = \Phi_T(\gamma_0)$, according to the time evolution $\Phi_T\colon M\to M$ over time $T$.
We may use different local coordinates to represent $\gamma_0$ and $\gamma_T$ as $d_\mathrm{E}$-vectors; we describe them as $\vect{\gamma}_0=(x_1, \dotsc, x_{2d-1})$ and $\vect{\gamma}_T=(x_1', \dotsc, x_{2d-1}')$.

The behaviour of the time evolution $\Phi_T$ near $\vect{\gamma}_0$ is linearly approximated by its Jacobian:
\begin{equation}
 \Phi_T(\vect{\gamma}_0 + \delta\vect{\gamma})
  \approx \vect{\gamma}_T + \mat{J}_T\cdot\delta\vect{\gamma}
  \qquad (\lVert\delta\vect{\gamma}\rVert\to 0),
  \label{LinearApprox}
\end{equation}
where $\mat{J}_T=\mat{J}_T(\vect{\gamma}_0)$ is a $(2d-1)\times(2d-1)$ matrix defined as
\begin{equation}
 [\mat{J}_T(\vect{\gamma}_0)]_{j,k} = \frac{\partial x_j'}{\partial x_k}
  .
\end{equation}

The singular values ${M}_1 \le \dotsb \le {M}_{2d-1}$ of the matrix $\mat{J}_T$ can be obtained as the eigenvalues of $(\mat{J}_T^{\rm t} \mat{J}_T)^{1/2}$, where $\rm t$ denotes the transposition.
By appropriately changing the coordinate systems $(x_1,\dotsc,x_{2d-1})$ and $(x_1', \dotsc, x_{2d-1}')$, we may regard $\mat{J}_T$ as a diagonal matrix with diagonal entries ${M}_1, \ldots, {M}_{2d-1}$, where \eqref{LinearApprox} can be rewritten as
\begin{equation}
 \Phi_T(x_1+\delta x_1,\dotsc, x_{2d-1}+\delta x_{2d-1})
 \approx (x_1'+{M}_1\delta x_1, \dotsc, x_{2d-1}'+{M}_{2d-1}\delta x_{2d-1}).
 \label{ScalingFactors}
\end{equation}
That is, the $j$-th component is rescaled by ${M}_j$ through the time evolution.

As $T$ goes to infinity, the singular values $M_j = M_j(\gamma_0,T)$ are governed by the Lyapunov exponents ${\Lambda}_1 \le \dotsb \le {\Lambda}_{2d-1}$ as
\begin{equation}
 M_j(\gamma_0,T) \sim e^{\Lambda_jT}. \label{LyapunovEstimates}
\end{equation}
According to Oseledets' theorem\cite{oseledets68}, in a volume-preserving dynamical system, the limit
\begin{equation}
 \mat{L} = \lim_{T\to\infty} (\mat{J}_T^{\rm t} \mat{J}_T)^{1/2T}
  \label{LyapunovLimit}
\end{equation}
exists for almost all $\gamma_0$.
The eigenvalues of $\mat{L}=\mat{L}(\gamma_0)$ are independent of the initial state $\gamma_0$, the logarithms of which give the Lyapunov exponents.

We may choose the coordinate system $(x_1,\dotsc,x_{2d-1})$ such that $\mat{L}$ becomes diagonal, with which the linear approximation \eqref{ScalingFactors} holds for $M_j \sim e^{\Lambda_j T}$.
Although the coordinate system $(x_1,\dotsc,x_{2d-1})$ is determined locally on $\gamma_0$, these local coordinates can be integrated to form one coordinate system\cite{hammerlindl11}.
In this coordinate system, the approximation in \eqref{ScalingFactors} is as good as the convergence of the limit \eqref{LyapunovLimit}.

We also note that a similar analysis on the backward time evolution $\Phi_T^{-1} = \Phi_{-T}$ leads to the reversed Lyapunov exponents $-\Lambda_{2d-1}\le \dotsb \le -\Lambda_1$.
The time reversal invariance of the Hamiltonian system requires $\Lambda_j = -\Lambda_{2d-j}$, and therefore the spectrum of Lyapunov exponents satisfies the following condition:
\begin{equation}
 -\lvert\Lambda_{1}\rvert \le \dotsb \le -\lvert\Lambda_{d-1}\rvert
  \le \Lambda_d=0 \le
  \lvert\Lambda_{d-1}\rvert \le \dotsb \le \lvert\Lambda_{1}\rvert.
  \label{SymmetricLyapunov}
\end{equation}

Let us now consider the time reversal test, where $\vect{\gamma}_T$ is perturbed into $\vect{\gamma}_T' = (x_1'+\delta x_1', \dotsc, x_{2d-1}'+\delta x_{2d-1}')$.
The noises $\delta x_1', \dotsc, \delta x_{2d-1}'$ are independently subject to Gaussian distributions of the mean $0$ and the standard deviation $l$.

We recall that the spread of $S_0$, the support of the initial state, is as large as $w$ in any direction.
To estimate the R\'enyi-0 divergence, we roughly approximate $S_0$ by a hyper-rectangular region as
\begin{equation}
 S_0 \sim \{ (y_1,\dotsc,y_{2d-1}) \mid a_j \le y_j \le b_j \}.
  \label{InitialBox}
\end{equation}
Here the interval $[a_j,b_j]$ contains $x_j$ and its width $b_j-a_j$ is comparable to $w$.
With the linear approximation in \eqref{ScalingFactors}, the support $S_T$ of the evolved state $\rho_T$ assumes the form of
\begin{equation}
 S_T \sim \{ (y_1',\dotsc,y_{2d-1}') \mid M_j a_j \le y_j' \le M_j b_j \}.
 \label{FinalBox}
\end{equation}
Therefore, the probability for $\gamma_T'$ to lie in $S_T$ can be estimated as
\begin{equation}
 \Prob[ \rho_T(\gamma_T')>0 \mid \gamma_0]
  \approx \prod_{j=1}^{2d-1} \frac{1}{\sqrt{2\pi}l}\int_{M_ja_j}^{M_jb_j}
  \exp \biggl[-\frac{(y_j'-x_j')^2}{2l^2}\biggr]dy_j'. \label{ProductIntegral}
\end{equation}
The behavior of integrals in \eqref{ProductIntegral} depends on the ratio of $M_jw$ to $l$.
When $M_jw \agt l$, this integral becomes of the order of unity, since the interval $[M_ja_j, M_jb_j]$ contains a significant part of the Gaussian peak.
When $M_jw\ll l$, the integral contains only a small fraction of the Gaussian peak, and therefore remains of the order of $M_jw/l$.
From these observations, we obtain
\begin{equation}
 \frac{1}{\sqrt{2\pi}l}\int_{M_ja_j}^{M_jb_j} \exp \biggl[-\frac{(y_j'-x_j')^2}{2l^2}\biggr]dy_j'
  \sim \min \{1,M_jw/l\}.
\end{equation}

Therefore, the conditional probability in \eqref{FinalBox} is estimated to be
\begin{equation}
 \Prob[ \rho_T(\gamma_T')>0 \mid \gamma_0]
  \sim \prod_{j=1}^{2d-1} \min \{1, M_jw/l \} = (w/l)^\alpha M_1\dotsm M_\alpha,
  \label{SpecificCondProb}
\end{equation}
where we denote by $\alpha$ the greatest integer that satisfies the scale separation $M_\alpha w/l \ll 1$.
Although this criterion for the scale separation can be ambiguous for finite $T$, the ambiguity disappears as $T$ goes to infinity.

Now we employ the Lyapunov estimates in \eqref{LyapunovEstimates} to obtain
\begin{equation}
 \Prob[ \rho_T(\gamma_T')>0 \mid \gamma_0]
  \sim (w/l)^\alpha\exp [(\Lambda_1+\cdots+\Lambda_\alpha)T].
\end{equation}
With sufficiently large $T$, the condition for the scale separation for $\alpha$ is independent of the initial state $\gamma_0$ and can be written as
\begin{equation}
  \Lambda_\alpha < \frac{1}{T}\ln \frac{l}{w} < \Lambda_{\alpha+1}.
   \label{ScalingSeparation}
\end{equation}
Therefore, we may remove the condition on $\gamma_0$ from the probability and finally obtain
\begin{equation}
 D_0 = -\ln \Prob[ \rho_T(\gamma_T')>0 ]
  \approx \alpha \ln l  - \alpha \ln w - (\Lambda_1 \dotsb + \Lambda_\alpha)T.
  \label{LinearRelation}
\end{equation}
Hence we have a linear relation between $D_0$ and $\ln l$ (and also $T$) in the regime of equation~\eqref{ScalingSeparation}.
In particular, when $T$ tends to infinity, the middle term of equation~\eqref{ScalingSeparation} tends to zero and therefore $\alpha$ is the number of negative Lyapunov exponents.
When $l<w$, this number is generally $d-1$ in a time-reversal-invariant system with Lyapunov exponents \eqref{SymmetricLyapunov}, unless some exponents other than $\Lambda_d$ happen to be zero.

\begin{figure}[tb]
  \centering
  \includegraphics[bb=0 0 340 376,height=0.3\textheight]{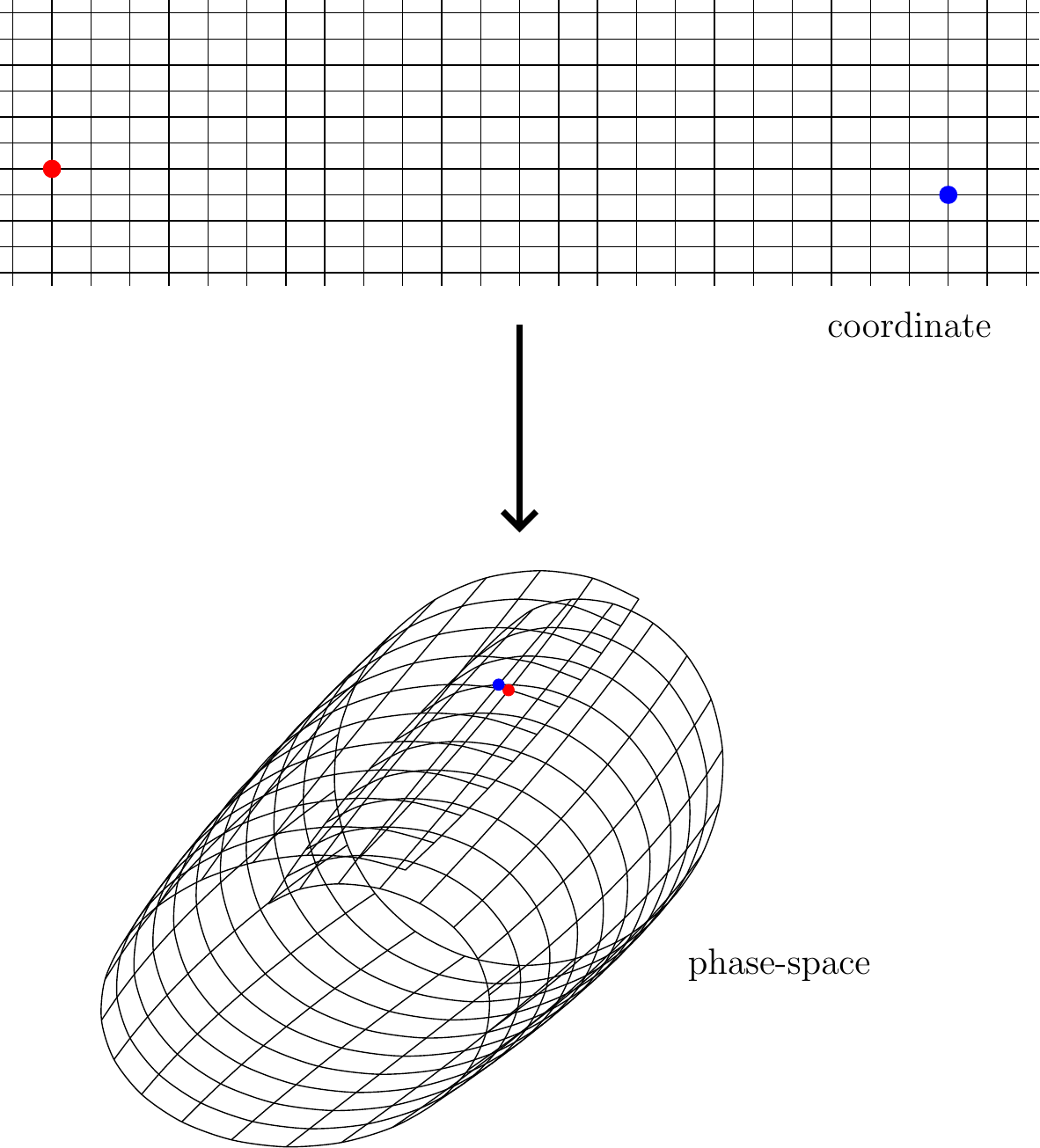}
  \caption{\label{FoldedCoord}
  Schematic picture of a coordinate in the phase space.
  The blue point appears far from the red point in this coordinate representation, although they are close to each other in the physical phase space.
  }
 \end{figure}

The linear relation~\eqref{LinearRelation} collapses for extremely large $T$, where the Poincar\'e recurrence occurs.  In the argument of this section, the long-time evolution affects the mapping from the phase space to the coordinate system $(x_1,\dotsc,x_{2d-1})$.
Although the perturbed state $\vect{\gamma}_T'$ and the unperturbed state $\vect{\gamma}_T$ look different from each other in the coordinate representation, it is possible that they are close in the phase space, which leads to wrong conclusions about whether $\vect{\gamma}_T'$ resides in the support of $\rho_T$ (See Figure~\ref{FoldedCoord}).
 This type of errors can emerge when the spread of $S_T$ is at least larger than the system size, but numerical results show that equation~\eqref{LinearRelation} holds beyond this time scale.
 A reasonable estimate for $T$ when equation~\eqref{LinearRelation} ceases to be valid is yet to be found.

\section{Details and precision of numerical simulations}

\subsection{Details of numerical simulations}
Throughout our numerical simulations, the geometry of the stadium is set to $R=1$ and $L=1$.
The velocity of the billiard is set to be unity.
The initial state $\rho_0$ is prepared to be a uniform distribution in the cube with the side $w$ and the least-valued vertex $(x_0,y_0,\theta_0)=(0.5,0.5,\pi/4)$.

In Fig.~1 in the main text, we set $w=0.01$ and $l=e^{-15}$.

In Fig.~2a in the main text, we set $w=0.01$ and calculate the R\'enyi-0 divergence $D_0$ by the Monte-Carlo method as described in the main text, while increasing the length scale $l$ from $e^{-20}$ to $1$ by the multiplication of $e^{0.2}$ and varying $T$ from $0$ to $70$ by the increment of $2$.
For each pair of $l$ and $T$, we sample $10^9$ events from a low-discrepancy sequence generated by the additive recurrence.
Although the numerical precision decreases as the value of $D_0$ increases, the relative statistical error can be estimated to be at most two percent.
In Fig.~2b, we calculate $dD_0/d\ln l$ by locally applying the least-square fitting to $D_0$ with respect to $\ln l$.

In Fig.~3a in the main text, we calculate the probability distribution of the empirical entropy production by setting $w=0.1$ and $l=e^{-10}$.
To evaluate the empirical entropy, we should calculate the probability distribution $\tilde \rho^l_0$ in the support of $\rho_0$.
To this aim, we divide the support of $\rho_0$ into $8^3$ sub-cubes.
In each sub-cube, we generate $2^{24}\sim 2\cdot 10^7$ samples as the initial state and perform the time reversal test.
The total sampling number is therefore $2^{33}\sim 9\cdot 10^9$.
By accumulating the events that terminate in each sub-cube, we numerically evaluate the probability distribution $\tilde \rho^l_0$.
We increase $T$ from $0$ to $70$ by an increment of $2$.
As $T$ increases, the statistical error increases because $\tilde\rho^l_0$ diffuses and the number of events that return to each sub-cube decreases.
Nevertheless, the relative statistical errors are at most five percent.
From the probability distribution of the empirical entropy production in Fig.~3a, the data in Fig.~3b, c, and d are calculated.
In Fig.~3e, we set $l=e^{-3}w$ and decrease $w$ from $e^{-1}$ to $e^{-5}$ by the multiplication of $e^{-1}$.
For each $w$ and $T$, the R\'enyi-0 divergence $D_0$ is evaluated from $10^{9}$ samples.

\subsection{Chaos and numerical precision}
\begin{figure}
	\flushleft\hspace{0.1\textwidth}{\bf a}
	\hspace{0.4\textwidth}{\bf b}\\
	\centering
	\includegraphics[width=0.8\textwidth]{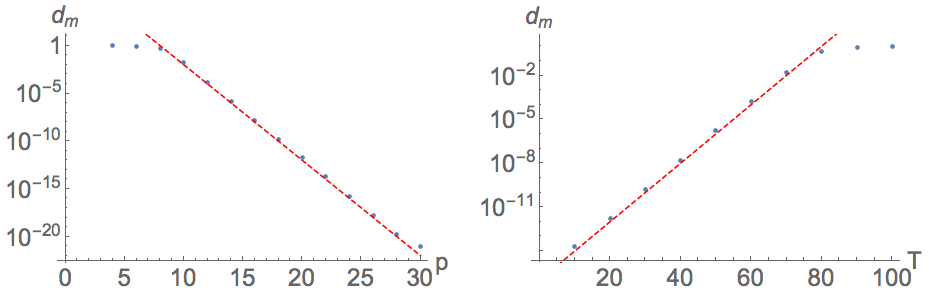}
	\caption{\label{fig:Precision}
		Errors caused by the numerical precision in the time reversal test with no added noises.
		{\bf a,} Dependence of the median of the distances $d_{\rm m}$ on the numerical precision $p$.
		By conducting the time reversal test with no noises, we obtain the distribution of the Euclidean distance between the initial state and the pullbacked state in phase space and calculate their median.
		The numerical precision is varied while the evolution time is fixed to be $T=40$.
		{\bf b,} Dependence of the median of the distances $d_{\rm m}$ on the evolution time $T$.
		The evolution time is varied with the numerical precision fixed to be $p=16$, i.e., the double precision.
		The red dashed line in each figure represents the estimate of errors given by equation~(\ref{Precision}). 
	}
\end{figure}

In this section, we argue that the double-precision arithmetics is enough to simulate our chaotic system under the present parameters.
To numerically examine the effect of a finite precision, we conduct the time reversal test with no noise.
Let $10^{-p}$ be the absolute precision of the floating number that we use in our simulations.
The largest error originates from the error in the stable direction at the final state.
When the velocity is reversed, the direction becomes unstable and the error grows exponentially as $e^{\Lambda T}$.
Therefore, the typical error is expected to be 
\begin{equation}\label{Precision}
	\epsilon\simeq 10^{-p} e^{\Lambda T} \simeq 10^{0.2 T-p},
\end{equation}
where we use $\Lambda=0.46$ for our Bunimovich billiard.
We numerically confirm this estimate by varying the precision of floating numbers as shown in Fig.~S\ref{fig:Precision}.
Hence, the reliability condition for our numerical simulation is $w\gg\epsilon\simeq 10^{0.2 T-p}$.
Therefore, for $w=0.01$ and the double precision ($p=16$), the reliability is guaranteed for $T\lesssim 70$, which covers the range of Fig.~2 in the main text.

\end{document}